\documentclass{article}

\usepackage{amsmath}
\usepackage{amsthm}

\newcommand{\vc}{\mathbf} 

\newcommand{\mc}{\mathcal}
\newcommand{\mr}{\mathrm}

\usepackage{float}
\usepackage{graphicx}
\usepackage[outdir=./]{epstopdf}
\usepackage{amsfonts}
\usepackage{siunitx}
\usepackage{amsthm}
\usepackage{amsmath}
\usepackage[font={footnotesize,it}]{caption}
\usepackage{wrapfig}
\usepackage{subcaption}
\usepackage[english]{babel}
\usepackage{float}
\usepackage{parskip}
\usepackage{xcolor}
\usepackage{tikz}
\usepackage{pgfplotstable}

\begin{document}
%
\title{Learning Schizophrenia Imaging Genetics Data Via Multiple Kernel Canonical Correlation Analysis }

\author{Owen Richfield\\
		Department of Biomedical Engineering\\
		Tulane University\\
		New Orleans, LA 70118, USA\\
		Email: orichfie@tulane.edu 
	\and 
	Md. Ashad Alam\\
	Department of Biomedical Engineering\\
		Tulane University\\
		New Orleans, LA 70118, USA\\
		Email: malam@tulane.edu
\and
Yu-Ping Wang\\
	Department of Biomedical Engineering\\
		Tulane University\\
		New Orleans, LA 70118, USA\\
		Email: wyp@tulane.edu
\and 
Vince Calhoun\\
Department of Electrical and \\Computer Engineering\\
		University of New Mexico and \\The Mind Research Network\\
		Albuquerque, NM 87131, USA\\
		Email: vcalhoun@unm.edu}


%


\maketitle

\begin{abstract}

Kernel and Multiple Kernel Canonical Correlation Analysis (CCA) are employed to classify schizophrenic and healthy patients based on their SNPs, DNA Methylation and fMRI data. Kernel and Multiple Kernel CCA are popular methods for finding nonlinear correlations between high-dimensional datasets. Data was gathered from 183 patients, 79 with schizophrenia and 104 healthy controls. Kernel and Multiple Kernel CCA represent new avenues for studying schizophrenia, because, to our knowledge, these methods have not been used on these data before. Classification is performed via k-means clustering on the kernel matrix outputs of the Kernel and Multiple Kernel CCA algorithm. Accuracies of the Kernel and Multiple Kernel CCA classification are compared to that of the regularized linear CCA algorithm classification, and are found to be significantly more accurate. Both algorithms demonstrate maximal accuracies when the combination of DNA methylation and fMRI data are used, and experience lower accuracies when the SNP data are incorporated.
\end{abstract}


%
\maketitle

\section{Introduction}
Schizophrenia is a complex neurological disorder whose manifestation can be attributed to numerous genetic, epigenetic and environmental factors. Genomic data have been used to identify genes at risk of causing schizophrenia while brain imaging techniques such as fMRI allow researchers to locate brain regions whose abnormal behaviors correlate to symptoms of the disorder. In the interest of leveraging data to produce more powerful conclusions on the causes of schizophrenia and to improve diagnosis of this and other complex mental disorders, the integration of these datasets via canonical correlation analysis (CCA) have been employed \cite{Dongdong-14}.

Linear or classical CCA seeks to obtain coefficients whose projections maximize the linear correlations between two or more (in the case of multiple CCA) sets of variables. Due to its versatility in integrating data, this method has been utilized by numerous scientists and researchers in the fields of statistics, biometrics, economics, and social science. Additionally, more robust methods of performing CCA have been developed since its inception \cite{Ashad-14T}.

Linear CCA assumes linearity of the data being studied, however in many cases it cannot be assumed that the two or more datasets are linearly correlated. Through the use of a Reproducing Kernel Hilbert Space (RKHS), kernel methods better maximize correlation between nonlinearly-correlated datasets and thus can be more attractive. The principle of Kernel and Multiple Kernel CCA, like classical linear CCA, is to find basis vectors by which the projection of two or more (in the case of Multiple Kernel CCA) datasets maximizes the correlation between these projections \cite{Ashad-14T}. In the case of Kernel and Multiple Kernel CCA, the datasets are projected into a Reproducing Kernel Hilbert Space, which can be of far higher dimension than the linear space into which classical CCA will project the data.

In this work, the reproducing kernel is used in Kernel and Multiple Kernel CCA to maximize nonlinear correlations between imaging genomics datasets. Kernel and Multiple Kernel CCA are employed in an algorithm to classify patients as ``healthy" or ``schizophrenic" based on their individual SNP, DNA Methylation and fMRI voxel information.

The rest of this work is organized as follows: Section 2 will include theory on linear CCA and multiple CCA; Section 3 will include theory on the Reproducing Kernel Hilbert Space, Kernel CCA and Multiple Kernel CCA; Section 4 will detail methods of experimentation and classification, including the data used in the experiments; and Section 5 will include conclusions and future directions for the research.

\section{Linear CCA and Multiple CCA}
In this section, we review the method of regularized linear canonical correlation analysis (CCA) of two or more (in the case of multiple CCA) datasets.
The goal of linear canonical correlation analysis is to determine the linear correlation between two or more datasets, and to develop coefficients that maximize this correlation. In mathematical terms, given $p$ sets of variables $[\mathbf{X}_1,\mathbf{X}_2,...,\mathbf{X}_p]$, linear CCA seeks to find vectors $\mathbf{w}_1,\mathbf{w}_2,...,\mathbf{w}_p$ such that correlation between the sets $\rho$, is maximized. This formulation is represented as follows:

\begin{equation}
\label{corr_mult}
\max_{\mathbf{w}_1,\mathbf{w}_2,...,\mathbf{w}_p} \rho = \frac{\sum_{j,k}\mathbf{w}_j^T\mathbf{C}_{j,k}\mathbf{w}_k}{\prod_{k=1}^p\sqrt{\mathbf{w}_k^T\mathbf{C}_{k,k}\mathbf{w}_k}},
\end{equation}

where $\mathbf{C}_{j,k}$ is the covariance matrix of sets $\mathbf{X}_j,\mathbf{X}_k$. Note that in the case that $p=2$, Equation \ref{corr_mult} becomes

\begin{equation}
\label{corr_2}
\max_{\mathbf{w}_1,\mathbf{w}_2} \rho = \frac{\mathbf{w}_1^T\mathbf{C}_{1,2}\mathbf{w}_2}{\sqrt{\mathbf{w}_1^T\mathbf{C}_{1,1}\mathbf{w}_1}\sqrt{\mathbf{w}_2^T\mathbf{C}_{2,2}\mathbf{w}_2}},
\end{equation}
the classical maximization problem to be solved by linear canonical correlation analysis. As such, multiple CCA can be viewed as a generalization of CCA that accepts more than two datasets.

To search for a maximum correlation, we solve the following eigenvalue problem:
\begin{align}
\begin{bmatrix} 0 & \mathbf{C}_{1,2} & \cdots & \mathbf{C}_{1,p} \\
\vdots & \vdots & \ddots & \vdots \\
\mathbf{C}_{p,1} & \mathbf{C}_{{p-1},1} & \cdots & 0  \end{bmatrix}
\left[ \begin{array}{c} \mathbf{w}_1 \\ \vdots \\ \mathbf{w}_p \end{array} \right]
= & \nonumber \\\rho
\begin{bmatrix} \mathbf{C}_{1,1} & \cdots & 0 \\ \vdots & \ddots & \vdots \\ 0 & \cdots & \mathbf{C}_{p,p} \end{bmatrix}
\left[ \begin{array}{c} \mathbf{w}_1 \\ \vdots \\ \mathbf{w}_p \end{array} \right].
\end{align}
Due to the possibility of singularity of the diagonal matrix in the specified eigenvalue problem \cite{Ashad-14T}, a regularization term is added as follows:

\begin{align}
\begin{bmatrix} 0 & \mathbf{C}_{1,2} & \cdots & \mathbf{C}_{1,p} \\
\vdots & \vdots & \ddots & \vdots \\
\mathbf{C}_{p,1} & \mathbf{C}_{{p-1},1} & \cdots & 0  \end{bmatrix}
\left[ \begin{array}{c} \mathbf{w}_1 \\ \vdots \\ \mathbf{w}_p \end{array} \right]
= & \nonumber \\\rho
\begin{bmatrix} \mathbf{C}_{1,1}+\lambda\mathbf{I} & \cdots & 0 \\ \vdots & \ddots & \vdots \\ 0 & \cdots & \mathbf{C}_{p,p}+\lambda\mathbf{I} \end{bmatrix}
\left[ \begin{array}{c} \mathbf{w}_1 \\ \vdots \\ \mathbf{w}_p \end{array} \right].
\end{align}
for $\lambda$ a small regularization parameter. With only two datasets ($p=2$), the multiple CCA becomes the regular linear CCA and the eigenvalue problem becomes

\begin{align}
\begin{bmatrix} 0 &\mathbf{C}_{1,2} \\
\mathbf{C}_{2,1} &  0  \end{bmatrix}
\left[ \begin{array}{c} \mathbf{w}_1 \\ \mathbf{w}_2 \end{array} \right]
= & \nonumber \\ \rho \begin{bmatrix} \mathbf{C}_{1,1}+\lambda\mathbf{I} & 0 \\ 0 & \mathbf{C}_{2,2}+\lambda\mathbf{I} \end{bmatrix}
\left[ \begin{array}{c} \mathbf{w}_1 \\ \mathbf{w}_2 \end{array} \right].\end{align}

\section{ Kernel CCA and multiple kernel CCA}
In this section, we review the single and Multiple Kernel Canonical Correlation Analysis (CCA).
\subsection{Kernel CCA}
\label{sec:CKCCA}
The aim of Kernel CCA is to seek two sets of functions in the RKHS for which the correlation (Corr) of random variables is maximized. Given two sets of random variables $X$ and $Y$ with two   functions in the RKHS, $f_{X}(\cdot)\in \mc{H}_X$  and  $f_{Y}(\cdot)\in \mc{H}_Y$, the optimization problem of the random variables $f_X(X)$ and $f_Y(Y)$ is
\begin{eqnarray}
\label{ckcca1}
\max_{\substack{f_{X}\in \mc{H}_X,f_{Y}\in \mc{H}_Y \\ f_{X}\ne 0,\,f_{Y}\ne 0}}\mr{Corr}(f_X(X),f_Y(Y)).
\end{eqnarray}
The optimizing functions $f_{X}(\cdot)$ and $f_{Y}(\cdotp)$ are determined up to scale.

Using a finite sample, we are able to estimate the desired functions. Given an i.i.d sample, $(X_i,Y_i)_{i=1}^n$ from a joint distribution $F_{XY}$, by taking the inner products with elements or ``parameters" in the RKHS, we have features
$f_X(\cdot)=\langle f_X, \Phi_X(X)\rangle_{\mc{H}_X}= \sum_{i=1}^na_X^ik_X(\cdot,X_i) $ and
$f_Y(\cdot)=\langle f_Y, \phi_Y(Y)\rangle_{\mc{H}_Y}=\sum_{i=1}^na_Y^ik_Y(\cdot,Y_i)$, where $k_X(\cdot, X)$ and $k_Y(\cdot, Y)$ are the associated kernel functions for $\mc{H}_X$ and $\mc{H}_Y$, respectively. The kernel Gram matrices are defined as   $\vc{K}_X:=(k_X(X_i,X_j))_{i,j=1}^n $ and $\vc{K}_:=(k_Y(Y_i,Y_j))_{i,j=1}^n $.  We need the centered kernel Gram matrices $\vc{M}_X=\vc{C}\vc{K}_X\vc{C}$ and $\vc{M}_Y=\vc{C}\vc{K}_Y\vc{C}$, where $\vc{C} = \vc{I}_n -\frac{1}{n}\vc{B}_n$ with $\vc{B}_n = \vc{1}_n\vc{1}^T_n$ and $\vc{1}_n$ is the vector with $n$ ones. The empirical estimate of Eq. (\ref{ckcca1}) is then given by
\begin{eqnarray}
\label{ckcca6}
\max_{\substack{f_{X}\in \mc{H}_X,f_{Y}\in \mc{H}_Y \\ f_{X}\ne 0,\,f_{Y}\ne 0}}\frac{\widehat{\rm{Cov}}(f_X(X),f_Y(Y))}{[\widehat{\rm{Var}}(f_X(X))]^{1/2}[\widehat{\rm{Var}}(f_Y(Y))]^{1/2}} \nonumber
\end{eqnarray}
where
\begin{align*}
& \widehat{\rm{Cov}}(f_X(X),f_Y(Y))
= \frac{1}{n} \vc{a}_X^T\vc{M}_X\vc{M}_Y \vc{a}_Y \\
& \widehat{\rm{Var}}( f_X(X))
=\frac{1}{n} \vc{a}_X^T\vc{M}_X^2 \vc{a}_X \,  \\ &\widehat{\rm{Var}}( f_Y(Y))=\frac{1}{n} \vc{a}_Y^T\vc{M}_Y^2 \vc{a}_Y,
\end{align*}
where $\vc{a}_{X}$ and $\vc{a}_{Y}$ are the directions of $X$ and $Y$, respectively. Solving the above maximization problem is then analogous to solving the eigenvalue problem in (\ref{ckcca7}):
\begin{align}
\label{ckcca7}
\begin{bmatrix}
0      & \vc{M}_1\vc{M}_2  \\
\vc{M}_2\vc{M}_1      & 0  \\
\end{bmatrix}
\begin{bmatrix} \vc{a}_{X}\\
\vc{a}_{Y}\\
\end{bmatrix} =& \nonumber\\ \rho\begin{bmatrix}
\vc{M}_1\vc{M}_1     & 0  \\
0     &   \vc{M}_2\vc{M}_2  \\
\end{bmatrix}
\begin{bmatrix} \vc{a}_{X}\\
\vc{a}_{Y}\\
\end{bmatrix}.  \end{align}

Unfortunately, the naive kernelization (\ref{ckcca7}) of CCA  is trivial and non-zero solutions of generalized eigenvalue problem are $\rho=\pm 1$ \cite{Ashad-10, Back-02}. To overcome this problem, we introduce small regularization terms in the denominator of the right hand side of (\ref{ckcca7}) as
\begin{align}
\label{ckcca8}
\begin{bmatrix}
0      & \vc{M}_1\vc{M}_2  \\
\vc{M}_2\vc{M}_1      & 0  \\
\end{bmatrix}
\begin{bmatrix} \vc{a}_{X}\\
\vc{a}_{Y}\\
\end{bmatrix} =& \nonumber\\ \rho\begin{bmatrix}
(\vc{M}_1+\kappa I)^2     & 0  \\
0     &   (\vc{M}_2+\kappa I)^2  \\
\end{bmatrix}
\begin{bmatrix} \vc{a}_{X}\\
\vc{a}_{Y}\\
\end{bmatrix}  \end{align}
where the small regularized coefficient is $\kappa >0$.

\subsection{Multiple kernel CCA}
\label{sec:MKCCA}
Multiple  kernel CCA  seeks  more than  two sets of  functions in the RKHSs for which the correlation (Corr) of  random variables  is maximized. Given $p$ sets of random variables $X_1,  \cdots X_p$   and $p$ functions in the RKHS, $f_{1}(\cdot)\in \mc{H}_1$,$\cdots$, $f_{p}(\cdot)\in \mc{H}_p$, the optimization problem of  the random variables $f_1(X_1)$, $\cdots$, $f_p(X_p)$ is

\begin{eqnarray}
\label{mkcca1}
\max_{\substack{f_{1}\in \mc{H}_{X_i},\cdots, f_{p}\in \mc{H}_{X_i} \\ f_{1}\ne 0,\, \cdots, f_{p}\ne 0}}\sum_{j=1, j^\prime > j}^p\mr{Corr}(f_j(X_j),f_j^\prime(X_j^\prime)).
\end{eqnarray}
Given an i.i.d sample, $(X_{i1},X_{i2}, \cdots, X_{ip})_{i=1}^n$ from a joint distribution $F_{X_1, \cdots, X_p}$, by taking the inner products with elements or ``parameters" in the RKHS, we have  features

\begin{eqnarray}
\label{mkcca2}
f_1(\cdot)=\langle f_1, \Phi_1(X_1)\rangle_{\mc{H}_1}= \sum_{i=1}^na_{i1}k_1(\cdot,X_i), \nonumber\\
\vdots,\nonumber\\
f_p(\cdot)=\langle f_p, \phi_p(X_p)\rangle_{\mc{H}_p}=\sum_{i=1}^na_{ip}k_p(\cdot,X{_ip}),
\end{eqnarray}

where $k_1(\cdot, X_1), \cdots,  k_p(\cdot, X_p)$ are the associated kernel functions for $\mc{H}_1, \cdots, \mc{H}_p$, respectively. The kernel Gram matrices are defined as   $\vc{K}_1:=(k_1(X_{i1},X_{i^\prime 1}))_{i,i^\prime=1}^n$, $\cdots$, $\vc{K}_p:=(k_1(X_{ip},X_{i^\prime p}))_{i,i^\prime=1}^n$. Similar to  Section \ref{sec:CKCCA}, using this kernel Gram matrices, the centered kernel Gram matrices are defined as  $\vc{M}_1=\vc{C}\vc{K}_1\vc{C}$, $\cdots$,  $\vc{M}_p=\vc{C}\vc{K}_p\vc{C}$, where $ \vc{C} = \vc{I}_n -\frac{1}{n}\vc{B}_n$ with  $\vc{B}_n = \vc{1}_n\vc{1}^T_n$ and $\vc{1}_n$ is the vector with $n$ ones. As in the two sets of data the empirical estimate of Eq. (\ref{mkcca1}) is obtained using the generalized eigenvalue problem, as given by
following problem:
\begin{align}
\label{mkcca3}
\begin{bmatrix}
0      & \vc{M}_1\vc{M}_2 &\vc{M}_1\vc{M}_3 & \dots & \vc{M}_1\vc{M}_p \\
\vc{M}_2\vc{M}_1      & 0 & \vc{M}_2\vc{M}_3 & \dots & \vc{M}_2\vc{M}_p \\
\hdotsfor{5} \\
\vc{M}_p\vc{M}_1     & \vc{M}_p\vc{M}_2  & \vc{M}_p\vc{M}_3  & \dots & 0
\end{bmatrix}
\begin{bmatrix} a_1\\
a_2\\
\hdotsfor{1}\\
a_p
\end{bmatrix} =& \nonumber\\ \rho\begin{bmatrix}
(\vc{M}_1+\kappa I)^2     & 0 &0 & \dots & 0 \\
0     &   (\vc{M}_2+\kappa I)^2  & 0& \dots & 0 \\
\hdotsfor{5} \\
0     & 0  & 0 & \dots &  (\vc{M}_p+\kappa I)^2
\end{bmatrix}
\begin{bmatrix} a_1\\
a_2\\
\hdotsfor{1}\\
a_p
\end{bmatrix}.  \end{align}

\section{Classification via Kernel CCA and Multiple Kernel CCA}
\label{Class KCCA}
To classify patients according to the binary phenotype, combinations of the datasets selected from SNPs, fMRI, and DNA methylation are integrated via the Kernel CCA and Multiple Kernel CCA algorithms. In all cases a dummy set with only classification information is used. The classification algorithm relies upon 10-fold cross-validation, in which 9 folds are used to train the classifier and the remaining fold is a test. The classification algorithm is as follows:
\begin{enumerate}
	\item Define datasets $\vc{X_1...X_n}$, $2<n<4$ to use for classification algorithm. Remove test sets $\vc{\tilde{X}_1...\tilde{X}_n}$ and obtain training sets which will be denoted by $\vc{X_1...X_n}$ for simplicity.
	\item Use Kernel CCA or Multiple Kernel CCA to obtain functional coefficient vectors $\vc{f_1...f_n}$ and construct sets of matrices $\vc{M=M_1f_1...M_nf_n}$ and $\vc{\tilde{M}=\tilde{M}_1f_1...\tilde{M}_nf_n}$ for $\vc{M_i}$ and $\vc{\tilde{M}_i}$ the centered Gram matrices for the $i^{th}$ training and testing set, respectively.
	\item Classify $\vc{\tilde{M}}$ via k-means trained on $\vc{M}$, and obtain error of classification by checking against actual phenotype labels.
	\item Perform algorithm in ten-fold cross-validation, averaging errors after ten folds.
\end{enumerate}

\begin{table*}[]
	\caption{Classification error of schizophrenia data using  Linear CCA, Kernel CCA and and Multiple Kernel CCA.}
	\label{table_error}
	\centering
	\begin{tabular}{| c | c | c | }
		\hline
		Dataset Combination & Kernel CCA \% error & CCA \% error \\
		\hline
		SNP & 45.3552 & 49.7268 \\
		\hline
		fMRI & 39.8907 & 42.6230 \\
		\hline
		DNA Methylation & 30.0546 & 46.9945 \\
		\hline
		SNP, fMRI & 35.5191 & 37.7049\\
		\hline
		DNA Methylation, fMRI & 27.3224 & 37.7049\\
		\hline
		SNP, DNA Methylation & 31.694 & 45.3552\\
		\hline
		SNP, DNA Methylation, fMRI & 30.6011 & 43.1694 \\
		\hline
	\end{tabular}
\end{table*}
\section{Experiments}
\subsection{Imaging Genomics Data}
\label{Sec:MCIC}
The Mind Clinical Imaging Consortium (MCIC) has collected three types of data (SNPs, fMRI and DNA methylation) from 208 subjects including $92$ schizophrenic patients (age: $34\pm 11$, $22$ females) and $116$ (age: $32\pm 11$, $44$ females) healthy controls. Without missing data, the number of subjects is $183$ ($79$ schizophrenia (SZ) patients and $104$ healthy controls)\cite{Dongdong-14}.

{\bf SNPs}: For each subject (SZ patients  and healthy controls) a blood sample was taken and DNA was extracted. All subject genes typing was performed at the Mind Research Network using the Illumina Infinium HumanOmni1- Quad assay covering  $1140419$ SNP loci.  To form the final genotype calls and to perform a series of standard quality control procedures bead studio and PLINK software packages were applied, respectively. Additionally, those SNPS that could not be mapped cleanly to genes using the Scandb gene mapping resource were thrown out. The final dataset spans $723404$ loci having $24272$ genes based on $183$ subjects (those without missing data). Genotypes   ``aa"  (non-minor allele), ``Aa" (one minor allele) and ``AA" (two minor alleles) were coded as $0$, $1$ and $2$for each  SNP, respectively \cite{Dongdong-14} \cite{Chen-12}.

{\bf fMRI}: Participants' fMRI data was collected during their block design motor response to auditory stimulation. State-of-the-art approaches use mainly Participants' feedback and experts' observations for this purpose. The aim was to continuously monitor the patients, acquiring images with parameters (TR=2000 ms, TE= 30ms, field of view=22cam, slice thickness=4mm, 1 mm skip, 27 slices, acquisition matrix $64\times 64$, flip angle =$90^{\circ}$) on a  Siemens3T Trio Scanner and 1.5 T Sonata with echo-planar imaging (EPI). Data were pre-processed with SPM5 software  and  were realigned spatially normalized and resliced to $3\times 3 \times 3$ mm.  It was  smoothed with  a  $10\times 10 \times 10$ $mm^3$ Gaussian kernel and analyzed by multiple regression considering the stimulus and their temporal derivatives plus an intercept term as repressors .  Finally the stimulus-on versus stimulus-off contrast images were extracted with $53\times 63 \times 46$ mission measurements, excluding voxels without measurements. $41236$ voxels were extracted from $116$ ROIs based on the aal brain atlas for analysis \cite{Dongdong-14}.

{\bf DNA methylation}:DNA methylation is one of the main epigenetic mechanisms to regulate gene expression. It appears to be involved in the development of schizophrenia. In this paper, we investigated $27481$ DNA methylation markers in blood from $79$ schizophrenia patients and $104$ healthy controls. Participants come from the MCIC, a collaborative effort of 4 research sites.  For  more details, site information and enrollment for schizophrenia patients and healthy controls are in \cite{Liu-13}. All participants' symptoms were evaluated by the Scale of the Assessment of Positive Symptoms and the Scale of the Assessment of Negative symptoms \cite{Andreasen-84}. DNA from blood samples was measured by the Illumina Infinium Methylation27 Assay. The methylation value is calculated by taking the ratio of the methylated probe intensity and the total probe intensity.

\subsection{Results}

All combinations of the three datasets (including only one of each dataset) are used for classification via the algorithm detailed in Section \ref{Class KCCA}. Table \ref{table_error} depicts \% error of Kernel CCA classification. Linear CCA is also used to classify the patients for the comparison of the two methods.

From these results it is evident that the Kernel and Multiple Kernel CCA classification algorithms are significantly more accurate than the Linear and Multiple CCA for this dataset. When using only one of the three datasets, both algorithms achieve lowest accuracy when using SNPs for classification, while the Kernel CCA has maximum accuracy using DNA methylation and the Linear CCA has maximum accuracy using fMRI. It can be easily seen that both Multiple Kernel CCA and Multiple CCA achieve global maximum accuracy by using DNA methylation and fMRI together, while adding SNPs to the mix increases error.This seems straightforward given that both methods experience global minimum accuracy using SNPs only.

For the Kernel and Multiple Kernel CCA classification algorithms, using two datasets appears to achieve higher classification accuracy than using a single dataset, except for the case of combining SNPs with DNA Methylation, in which the SNP data appears to throw off the classification. Again, this loss of accuracy due to incorporation of the SNP data comes as little surprise given the low accuracy of the classifier when using SNPs alone. Overall, the results conclude that classification via Kernel and Multiple Kernel CCA achieves highest accuracy with the use of fMRI and DNA Methylation combined, and that the best dataset to use alone is DNA Methylation. Adding SNPs to the sets used for classification will likely decrease the accuracy. The highest accuracy achieved by the classifier is approximately 72.6\%, with Kernel CCA using the combination of fMRI and DNA methylation data.

\section{Conclusion and Future Work}
It is apparent that for classification of patients in a binary phenotype of ``healthy" versus ``schizophrenic," Kernel and Multiple Kernel CCA far surpasses linear CCA in accuracy. Both methods achieve maximal accuracy utilizing the combination of DNA methylation and fMRI, and achieve minimal accuracy classifying only on SNPs. From these results it is possible to conclude that in some cases the integration of multiple data modalities may yield higher accuracies in classification of complex neurological diseases such as schizophrenia. In this case, the combination of imaging and epigenetic factors produce better results than the incorporation of SNP variations. 

It also appears that in this case the nonlinear correlations between datasets produce more easily-separable and thus classifiable components than linear CCA, as evidenced by the Kernel  and Multiple Kernel CCA's superior accuracy to linear CCA. Both CCA and Kernel CCA serve as a feature extraction tool, based on which the classifier is used to separate patients from healthy controls. It appears Kernel and Multiple Kernel CCA can better reveal the relationship of three datasets and the best combination is fMRI and methylation. This work also demonstrates that the projection coefficients on the variants can serve as a distinct feature for classification.

For future work, parameter optimization must be employed to perfect results of the classification algorithm. Such parameters include kernel type, $k$ value for k-means clustering, and number of components that were used in the Kernel and Multiple Kernel CCA step. Actual changes may be made to the classification algorithm, such as using SVM instead of k-means classification. Given the complex nature of schizophrenia as a neurological disorder, the phenotype is complex and multidimensional. As such, in future work patients will be classified using a higher-dimensional phenotype space.
\section*{Acknowledgments}
The authors wish to thank the NIH (R01 GM109068, R01 MH104680, R01 MH107354) and NSF (1539067) for support.

\end{document}